\documentclass[12pt,onecolumn]{IEEEtran}

\usepackage{color}
\usepackage{amsmath}
\usepackage{amssymb}
\usepackage[dvips]{graphicx}
\usepackage{setspace}

\renewcommand{\baselinestretch}{1.82}

\newtheorem{lemma:coded_AWGN}{Lemma}[section]
\newtheorem{lemma:uncoded_sym}[lemma:coded_AWGN]{Lemma}
\newtheorem{lemma:uncoded_chip}[lemma:coded_AWGN]{Lemma}

\newtheorem{lemma:IFI}{Lemma}[section]
\newtheorem{lemma:MAI_MP}[lemma:IFI]{Lemma}

\begin{document}

\title{The Trade-off between Processing Gains of an Impulse Radio UWB System in the Presence of Timing Jitter$^{\textrm{\small{1}}}$}
\author{Sinan Gezici$^{2}$, \textit{Member, IEEE}, Andreas F. Molisch$^{3}$,
\textit{Fellow, IEEE},\\H. Vincent Poor$^{4}$, \textit{Fellow,
IEEE}, and Hisashi Kobayashi$^{4}$, \textit{Life Fellow, IEEE}}

\footnotetext[1]{This research is supported in part by the National
Science Foundation under grant CCR-99-79361, and in part by the New
Jersey Center for Wireless Telecommunications. Part of this work was
presented at the IEEE International Conference on Communications
2004.} \footnotetext[2]{Department of Electrical and Electronics
Engineering, Bilkent University, Bilkent, Ankara TR-06800, Turkey,
e-mail: gezici@ee.bilkent.edu.tr, Tel: +90(312) 290-3139, Fax:
+90(312) 266-4192} \footnotetext[3]{Mitsubishi Electric Research
Labs, 201 Broadway, Cambridge, MA 02139, USA, and Department of
Electroscience, Lund University, Box 118, SE-221 00 Lund, Sweden,
e-mail: Andreas.Molisch@ieee.org} \footnotetext[4]{Department of
Electrical Engineering, Princeton University, Princeton, NJ 08544,
USA, Tel: (609) 258-1816, Fax: (609) 258-2158, email:
\{hisashi,poor\}@princeton.edu}

\maketitle

\begin{abstract}
In time hopping impulse radio, $N_f$ pulses of duration $T_c$ are
transmitted for each information symbol. This gives rise to two
types of processing gain: (i) pulse combining gain, which is a
factor $N_f$, and (ii) pulse spreading gain, which is $N_c=T_f/T_c$,
where $T_f$ is the mean interval between two subsequent pulses. This
paper investigates the trade-off between these two types of
processing gain in the presence of timing jitter. First, an additive
white Gaussian noise (AWGN) channel is considered and approximate
closed form expressions for bit error probability are derived for
impulse radio systems with and without pulse-based polarity
randomization. Both symbol-synchronous and chip-synchronous
scenarios are considered. The effects of multiple-access
interference and timing jitter on the selection of optimal system
parameters are explained through theoretical analysis. Finally, a
multipath scenario is considered and the trade-off between
processing gains of a synchronous impulse radio system with
pulse-based polarity randomization is analyzed. The effects of the
timing jitter, multiple-access interference and inter-frame
interference are investigated. Simulation studies support the
theoretical results.

\end{abstract}

\newcounter{fnno}
\setcounter{fnno}{5}

\section{Introduction}

Recently, communication systems that employ ultra-wideband (UWB)
signals have drawn considerable attention. UWB systems occupy a
bandwidth larger than 500 MHz, and they can coexist with incumbent
systems in the same frequency range due to large spreading factors
and low power spectral densities. Recent Federal Communications
Commission (FCC) rulings \cite{FCC1}, \cite{FCC2} specify the
regulations for UWB systems in the US.

Commonly, impulse radio (IR) systems, which transmit very short
pulses with a low duty cycle, are employed to implement UWB systems
\cite{scholtz}-\cite{Win_2000_UWBforMA}. Although the short duration
of UWB pulses is advantageous for precise positioning applications
\cite{Sinan_magazin}, it also presents practical difficulties such
as synchronization, which requires efficient search strategies
\cite{Win_2007_Acq}. In an IR system, a train of pulses is sent and
information is usually conveyed by the positions or the amplitudes
of the pulses, which correspond to pulse position modulation (PPM)
and pulse amplitude modulation (PAM), respectively. Also, in order
to prevent catastrophic collisions among different users and thus
provide robustness against multiple access interference, each
information symbol is represented not by one pulse but by a sequence
of pulses and the locations of the pulses within the sequence are
determined by a pseudo-random time-hopping (TH) sequence
\cite{scholtz}.

\begin{figure}
\begin{center}
\includegraphics[width = 1\textwidth]{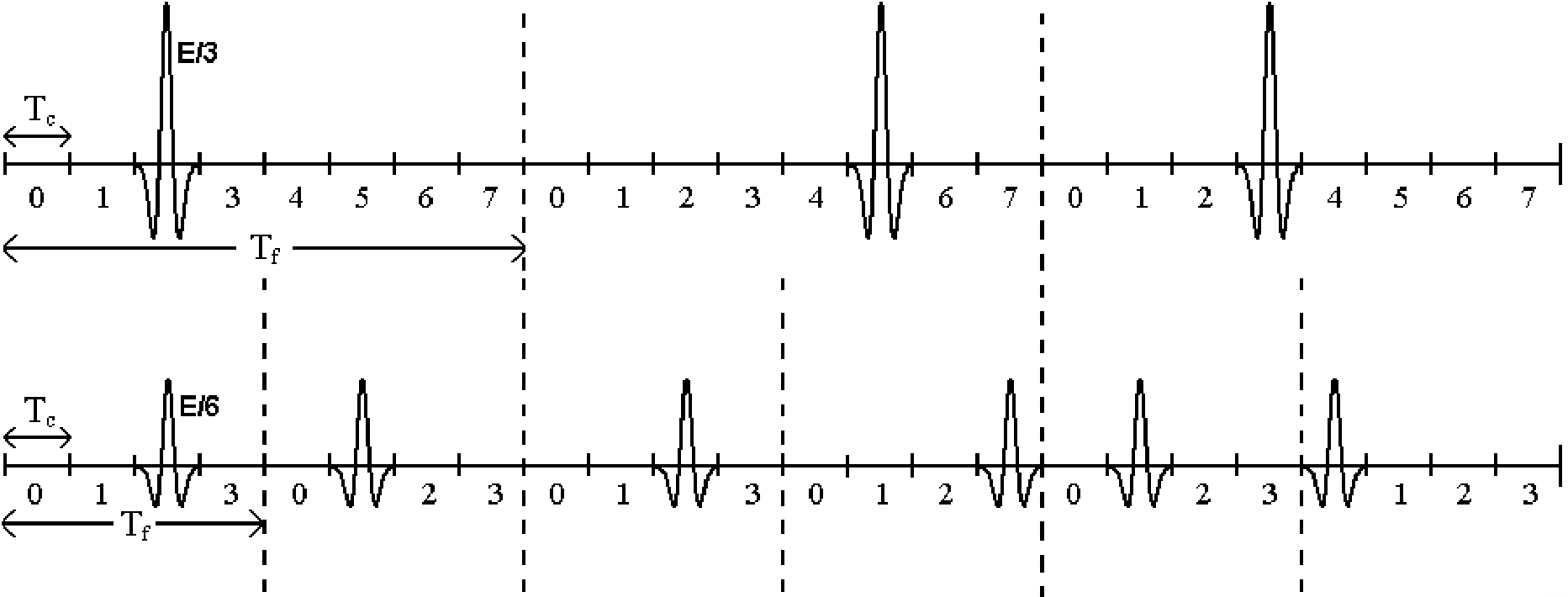}
\caption{Two different cases for a BPSK-modulated TH-IR system
without pulse-based polarity randomization when $N=24$. For the
first case, $N_c=8$, $N_f=3$, and the pulse energy is $E/3$. For
the second case, $N_c=4$, $N_f=6$, and the pulse energy is $E/6$.}
\label{fig:tradeOff}
\end{center}
\end{figure}

The number of pulses that are sent for each information symbol is
denoted by $N_f$. This first type of processing gain is called the
\textit{pulse combining gain}. The second type of processing gain
$N_c$ is the \textit{pulse spreading gain}, and is defined as the
ratio of average time between the two consecutive transmissions
($T_f$) and the actual transmission time ($T_c$); that is,
$N_c=T_f/T_c$. The total processing gain is defined as $N=N_cN_f$
and assumed to be fixed and large \cite{eran1}. The aim of this
paper is to investigate the trade-off between the two types of
processing gain, $N_c$ and $N_f$, and to calculate the optimal $N_c$
($N_f$) value such that bit error probability (BEP) of the system is
minimized$^{5}\footnotetext[5]{The FCC regulations also impose
restriction on peak-to-average ratio (PAR), which is not considered
in this paper \cite{FCC2}.}$. In other words, the problem is to
decide whether or not sending more pulses each with less energy is
more desirable in terms of BEP performance than sending fewer pulses
each with more energy (Figure \ref{fig:tradeOff}).


This problem is originally investigated in \cite{eran1}. Also
\cite{andy} analyzed the problem from an information theoretic point
of view for the single-user case. In \cite{eran1}, it is concluded
that in multiuser flat fading channels, the system performance is
independent of the pulse combining gain for an IR system with
pulse-based polarity randomization and it is in favor of small pulse
combining gain for an IR system without pulse-based polarity
randomization. However, the analysis is performed in the absence of
any timing jitter. Due to the high time resolution of UWB signals,
effects of timing jitter are usually not negligible
\cite{Ismail_jitter}-\cite{Lovelace_2002_Jitter UWB} in IR-UWB
systems. As will be observed in this paper, presence of timing
jitter has an effect on the trade-off between the processing gains,
which can modify the dependency of the BEP expressions on the
processing gain parameters. In this paper, the trade-off between the
two types of processing gain is investigated in the presence of
timing jitter for TH IR systems. First, transmission over an
additive white Gaussian noise (AWGN) channel is considered and the
trade-off is investigated for IR systems with and without
pulse-based polarity randomization. Both symbol-synchronous and
chip-synchronous cases are investigated. Also frequency-selective
channels are considered and the performance of a downlink IR system
with pulse-based polarity randomization is analyzed.

\newcounter{sec}
\setcounter{sec}{2} The remainder of the paper is organized as
follows. Section \Roman{sec} describes the signal model for an IR
system. Section \setcounter{sec}{3}\Roman{sec} investigates the
trade-off between processing gain parameters for IR systems with
and without pulse-based polarity randomization over AWGN channels.
In each case, the results for symbol-synchronous and
chip-synchronous systems are presented. Section
\setcounter{sec}{4}\Roman{sec} considers transmission over
frequency-selective channels and adopts a quite general Rake
receiver structure at the receiver. After the simulation studies
in Section \setcounter{sec}{5}\Roman{sec}, some conclusions are
made in Section \setcounter{sec}{6}\Roman{sec}.

\section{Signal Model}

Consider a BPSK random TH IR system where the transmitted signal
from user $k$ in an $N_u$-user setting is represented by the
following model:
\begin{gather}\label{eq:tran1}
s^{(k)}_{{\rm{tx}}}(t)=\sqrt{\frac{E_k}{N_f}}\sum_{j=-\infty}^{\infty}d^{(k)}_j\,
b^{(k)}_{\lfloor
j/N_f\rfloor}w_{{\rm{tx}}}(t-jT_f-c^{(k)}_jT_c-\epsilon_j^{(k)}),
\end{gather}
where $w_{{\rm{tx}}}(t)$ is the transmitted UWB pulse, $E_k$ is the
bit energy of user $k$, $\epsilon_j^{(k)}$ is the timing jitter at
$j$th pulse of the $k$th user, $T_f$ is the average time between two
consecutive pulses (also called the ``frame'' time), $T_c$ is the
pulse interval, $N_f$ is the number of pulses representing one
information symbol, which is called the pulse combining gain, and
$b^{(k)}_{\lfloor j/N_f \rfloor}\in \{+1,-1\}$ is the information
symbol transmitted by user $k$. In order to allow the channel to be
shared by many users and avoid catastrophic collisions, a random TH
sequence $\{c^{(k)}_j\}$ is assigned to each user, where $c^{(k)}_j
\in \{0,1,...,N_c-1\}$ with equal probability, and $c^{(k)}_j$ and
$c^{(l)}_i$ are independent for $(k,j)\ne (l,i)$. This TH sequence
provides an additional time shift of $c^{(k)}_jT_c$ seconds to the
$j$th pulse of the $k$th user. Without loss of generality,
$T_f=N_cT_c$ is assumed throughout the paper.

Two different IR systems are considered depending on $d^{(k)}_j$.
For IR systems with pulse-based polarity randomization \cite{paul},
$d^{(k)}_j$ are binary random variables taking values $\pm1$ with
equal probability and are independent for $(k,j)\ne (l,i)$.
Complying with the terminology established in \cite{eran1}, such
systems will be called ``coded" throughout the paper. The systems
with $d^{(k)}_j=1$, $\forall k,j$ are called ``uncoded''. This
second type of system is the original proposal for transmission over
UWB channels (\cite{scholtz}, \cite{gian}) while a version of the
first type is proposed in \cite{sadler}.

The timing jitter $\epsilon_j^{(k)}$ in (\ref{eq:tran1}) mainly
represents the inaccuracies of the local pulse generators at the
transmitters and is modeled as independent and identically
distributed (i.i.d.) among the pulses of a given user
\cite{2006_Kokalis_MUI jitter}, \cite{2005_Zhang_Jitter resist}.
That is, $\epsilon^{(k)}_j$ for $j=\ldots,-1,0,1,\ldots$ form an
i.i.d. sequence. Also the jitter is assumed to be smaller than the
pulse duration $T_c$, that is,
$\underset{j,k}{\max}|\epsilon_j^{(k)}|<T_c$, which is usually the
case for practical situations.

$N=N_cN_f$ is defined to be the total processing gain of the
system. Assuming a large and constant $N$ value \cite{eran1}, the
aim is to obtain the optimal $N_c$ ($N_f$) value that minimizes
the BEP of the system.

\section{AWGN Channels}

The received signal over an AWGN channel in an $N_u$-user system
can be expressed as
\begin{gather}\label{eq:rec_Flat}
r(t)=\sum_{k=1}^{N_u}\sqrt{\frac{E_k}{N_f}}\sum_{j=-\infty}^{\infty}d^{(k)}_j\,
b^{(k)}_{\lfloor
j/N_f\rfloor}w_{{\rm{rx}}}(t-jT_f-c^{(k)}_jT_c-\epsilon_j^{(k)}-\tau^{(k)})+\sigma_nn(t),
\end{gather}
where $w_{{\rm{rx}}}(t)$ is the received unit-energy UWB pulse,
$\tau^{(k)}$ is the delay of user $k$ and $n(t)$ is white Gaussian
noise with zero mean and unit spectral density.

Considering a matched filter (MF) receiver, the template signal at
the receiver can be expressed as follows, for the $i$th
information symbol:
\begin{gather}\label{eq:temp}
s^{(1)}_{{\rm{temp}}}(t)=\sum_{j=iN_f}^{(i+1)N_f-1}d_j^{(1)}w_{{\rm{rx}}}(t-jT_f-c^{(1)}_jT_c-\tau^{(1)}),
\end{gather}
where, without loss of generality, user $1$ is assumed to be the
user of interest. Also note that no timing jitter is considered
for the template signal since the jitter model in the received
signal can be considered to account for that jitter as well,
without loss of generality.

From (\ref{eq:rec_Flat}) and (\ref{eq:temp}), the MF output for user
$1$ can be expressed as follows$^{6,7}$\footnotetext[6]{The
self-interference term due to timing jitter is ignored since it
becomes negligible for large $N_c$ and/or small
$\textrm{E}\{\phi^2_w(T_c-|\epsilon^{(1)}|)\}$ values, where
$\phi_w(x)=\int_{-\infty}^{\infty}w_{{\rm{rx}}}(t)w_{{\rm{rx}}}(t-x)dt$.
However, it will be considered for the multipath case in Section
\ref{sec:MP}.} \footnotetext[7]{Subscripts for user and symbol
indices are omitted for $y$, $a$ and $n$ for simplicity.}:
\begin{gather}\label{eq:MF_out}
y=\int
r(t)s_{{\rm{temp}}}^{(1)}(t)dt\approx\sqrt{\frac{E_1}{N_f}}\,b_i^{(1)}\sum_{j=iN_f}^{(i+1)N_f-1}\phi_w(\epsilon_j^{(1)})+a+n,
\end{gather}
where the first term is the desired signal part of the output with
$\phi_w(x)=\int_{-\infty}^{\infty}w_{{\rm{rx}}}(t)w_{{\rm{rx}}}(t-x)dt$
being the autocorrelation function of the UWB pulse, $a$ is the
multiple-access interference (MAI) due to other users and $n$ is
the output noise, which is approximately distributed as $n\sim
\mathcal{N}(0,\,N_f\sigma_n^2)$.

The MAI term can be expressed as the sum of interference terms
from each user, that is,
$a=\sum_{k=2}^{N_u}\sqrt{\frac{E_k}{N_f}}a^{(k)}$, where each
interference term is in turn the summation of interference to one
pulse of the template signal:
\begin{gather}\label{eq:MAI_k}
a^{(k)}=\sum_{m=iN_f}^{(i+1)N_f-1}a_m^{(k)},
\end{gather}
where
\begin{gather}\label{eq:MAI_l}
a_m^{(k)}=d_m^{(1)}\int
w_{{\rm{rx}}}(t-mT_f-c_m^{(1)}T_c-\tau^{(1)})\sum_{j=-\infty}^{\infty}d_j^{(k)}b^{(k)}_{\lfloor
j/N_f\rfloor}w_{{\rm{rx}}}(t-jT_f-c^{(k)}_jT_c-\epsilon_j^{(k)}-\tau^{(k)})\,dt.
\end{gather}

As can be seen from (\ref{eq:MAI_l}), $a_m^{(k)}$ denotes the
interference from user $k$ to the $m$th pulse of the template
signal.

In this study, we consider chip-synchronous and symbol-synchronous
situations for the simplicity of the expressions. However, the
current study can be extended to asynchronous systems as well
\cite{sinan_T_SP}. In this paper, we will see that for coded
systems, the effect of the MAI is the same whether the users are
symbol-synchronous or chip-synchronous. However, for uncoded
systems, the average power of the MAI is larger, hence the BEP is
higher, when the users are symbol-synchronous.

\begin{figure}
\begin{center}
\includegraphics[width = 1\textwidth]{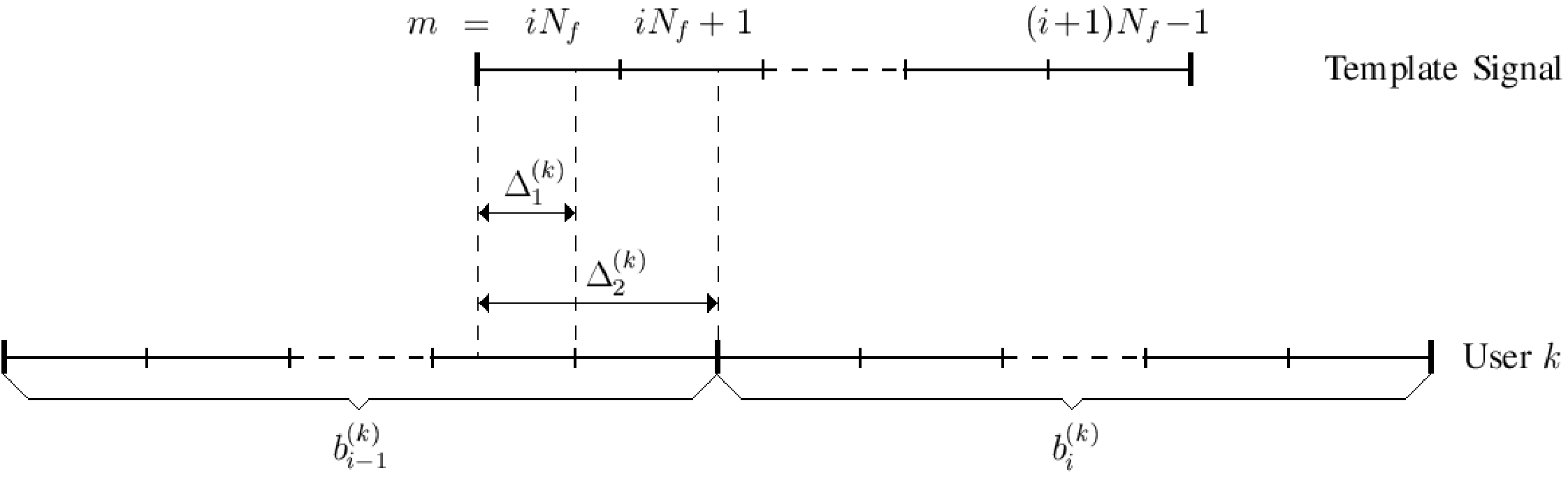}
\caption{The positions of the template signal and the signal of
user $k$.} \label{fig:Async}
\end{center}
\end{figure}

We assume, without loss of generality, that the delay of the first
user, $\tau^{(1)}$, is zero. Then, $\tau^{(k)}=0$ $\forall k$ for
symbol-synchronous systems. For chip-synchronous systems,
$\tau^{(k)}=\Delta_2^{(k)}T_c$, where
$\Delta_2^{(k)}\in\{0,1,\ldots,N-1\}$ with equal probability. Also
let $\Delta^{(k)}_1$ be the offset between the frames of user $1$
and $k$. Then, $\Delta^{(k)}_1=\mod\{\Delta_2^{(k)},N_c\}$ and
obviously, $\Delta_1^{(k)}\in\{0,1,...,N_c-1\}$ with equal
probability (Figure \ref{fig:Async}).


\subsection{Coded Systems}

For symbol-synchronous and chip-synchronous coded systems, the
following lemma approximates the probability distribution of
$a^{(k)}$ in (\ref{eq:MAI_k}):

\begin{lemma:coded_AWGN}\label{lem:coded_AWGN}
As $N\longrightarrow \infty$ and $\frac{N_f}{N_c}\longrightarrow
c>0$, $a^{(k)}$ is asymptotically normally distributed as
\begin{gather}\label{eq:flat_MAI_k}
a^{(k)} \sim {\mathcal{N}}(0\,,\,\gamma^{(k)}_2N_f/N_c),
\end{gather}
where
$\gamma^{(k)}_2=\textrm{E}\{\phi_w^2(\epsilon^{(k)})\}+\textrm{E}\{\phi_w^2(T_c-|\epsilon^{(k)}|)\}$.
\end{lemma:coded_AWGN}

\textit{Proof:} See Appendix \ref{app:coded_AWGN}.

From the previous lemma, it is observed that the distribution of
the MAI is the same whether there is symbol-synchronization or
chip-synchronization among the users, which is due to the use of
random polarity codes in each frame.

From (\ref{eq:MF_out}) and (\ref{eq:flat_MAI_k}), the BEP of the
coded IR system conditioned on the timing jitter of user $1$ can
be approximated as follows:
\begin{gather}\label{eq:cond_BEP_coded}
P_{e|\boldsymbol{\epsilon}_i^{(1)}}\approx
Q\left(\frac{\sqrt{\frac{E_1}{N_f}}
\sum_{j=iN_f}^{(i+1)N_f-1}\phi_w(\epsilon_j^{(1)})}
{\sqrt{\frac{1}{N_c}\sum_{k=2}^{N_u}E_k\gamma^{(k)}_2+N_f\sigma_n^2}}\right),
\end{gather}
where
$\boldsymbol{\epsilon}_i^{(1)}=[\epsilon^{(1)}_{iN_f}\ldots\epsilon^{(1)}_{(i+1)N_f-1}]$.

For large values of $N_f$, it follows from the Central Limit
Theorem (CLT) that
$\frac{1}{\sqrt{N_f}}\sum_{j=iN_f}^{(i+1)N_f-1}[\phi_w(\epsilon_j^{(1)})-\textrm{E}\{\phi_w(\epsilon_j^{(1)})\}]$
is approximately Gaussian. Then, using the relation
$\textrm{E}\{Q(X)\}=Q\left(\frac{\hat{\mu}}{\sqrt{1+\hat{\sigma}^2}}\right)$
for $X\sim\mathcal{N}(\hat{\mu},\hat{\sigma}^2)$ \cite{verdu}, the
unconditional BEP can be expressed approximately as follows:
\begin{gather}\label{eq:BEP_coded}
P_{e}\approx
Q\left(\frac{\sqrt{E_1}\mu}{\sqrt{\frac{E_1\sigma^2}{N_f}
+\frac{1}{N}\sum_{k=2}^{N_u}E_k\gamma^{(k)}_2+\sigma_n^2}}\right),
\end{gather}
where $\mu=\textrm{E}\{\phi_w(\epsilon_j^{(1)})\}$ and
$\sigma^2=\textrm{Var}\{\phi_w(\epsilon_j^{(1)})\}$.

From (\ref{eq:BEP_coded}), it is observed that the BEP decreases
as $N_f$ increases, if the first term in the denominator is
significant. In other words, the BEP gets smaller for larger
number of pulses per information symbol. We observe from
(\ref{eq:BEP_coded}) that the second term in the denominator,
which is due to the MAI, depends on $N_c$ and $N_f$ only through
their product $N=N_cN_f$. Therefore, the MAI has no effect on the
trade-off between processing gains for a fixed total processing
gain $N$. The only term that depends on how to distribute $N$
between $N_c$ and $N_f$ is the first term in the denominator,
which reflects the effect of timing jitter. This effect is
mitigated by choosing small $N_c$, or large $N_f$, which means
sending more pulses per information bit. Therefore, for a coded
system, keeping $N_f$ large can help reduce the BEP. Also note
that in the absence of timing jitter, (\ref{eq:BEP_coded}) reduces
to $P_{e}\approx
Q\left(\frac{\sqrt{E_1}}{\sqrt{\frac{1}{N}\sum_{k=2}^{N_u}E_k+\sigma_n^2}}\right)$,
in which case there is no effect of processing gain parameters on
BEP performance, as stated in \cite{eran1}.

\subsection{Uncoded Systems}

For coded systems, we have observed that the system performance is
the same for symbol-synchronous and chip-synchronous scenarios.
For an uncoded system, the effect of MAI changes depending on the
type of synchronism, as we study in this section.

First consider a symbol-synchronous system; that is,
$\tau^{(k)}=0$ $\forall k$ in (\ref{eq:rec_Flat}). In this case,
following lemma approximates the probability distribution of
$a^{(k)}$ in (\ref{eq:MAI_k}) for an uncoded system:
\begin{lemma:uncoded_sym}\label{lem:uncoded_sym}
As $N\longrightarrow \infty$ and $\frac{N_f}{N_c}\longrightarrow
c>0$, $a^{(k)}$ conditioned on the information bit $b_i^{(k)}$, is
approximately distributed as
\begin{gather}\label{eq:lem_uncod_sym}
a^{(k)}|b_i^{(k)} \sim
{\mathcal{N}}\left(\frac{N_f}{N_c}b_i^{(k)}\gamma^{(k)}_1\,,\,\frac{N_f}{N_c}\left[\gamma^{(k)}_2-\frac{(\gamma^{(k)}_1)^2}{N_c}+\frac{\beta^{(k)}_1}{N_c^2}+\frac{\beta^{(k)}_2}{N_c^3}\right]\right),
\end{gather}
where
\begin{align}\label{eq:parameters}\nonumber
\gamma^{(k)}_1&=\textrm{E}\{\phi_w(\epsilon^{(k)})\}+\textrm{E}\{\phi_w(T_c-|\epsilon^{(k)}|)\},\quad
\gamma^{(k)}_2=\textrm{E}\{\phi_w^2(\epsilon^{(k)})\}+\textrm{E}\{\phi_w^2(T_c-|\epsilon^{(k)}|)\},\\\nonumber
\beta^{(k)}_1&=2\textrm{E}\{\phi_w(T_c-|\epsilon^{(k)}|)\phi_w(\epsilon^{(k)})\}-2(\textrm{E}\{\phi_w(T_c-|\epsilon^{(k)}|)\})^2\\\nonumber
&+4\int_{-\infty}^{0}\phi_w(T_c+\epsilon^{(k)})p(\epsilon^{(k)})d\epsilon^{(k)}
\int_{0}^{\infty}\phi_w(T_c-\epsilon^{(k)})p(\epsilon^{(k)})d\epsilon^{(k)},\\
\beta^{(k)}_2&=2(\textrm{E}\{\phi_w(T_c-|\epsilon^{(k)}|)\})^2.
\end{align}
\end{lemma:uncoded_sym}

\textit{Proof:} See Appendix \ref{app:uncoded_sym}.

Note that for systems with large $N_c$, the distribution of
$a^{(k)}$ given the information symbol $b_i^{(k)}$ can be
approximately expressed as $a^{(k)}|b_i^{(k)} \sim
{\mathcal{N}}\left(b_i^{(k)}\gamma^{(k)}_1N_f/N_c\,,\,\frac{N_f}{N_c}[\gamma^{(k)}_2-(\gamma_1^{(k)})^2/N_c]\right)$.

First consider a two-user system. For equiprobable information
symbols $\pm1$, the BEP conditioned on timing jitter of the first
user can be shown to be
\begin{gather}
\textrm{P}_{e|\boldsymbol{\epsilon}^{(1)}}\approx\frac{1}{2}\,Q\left(\frac{\frac{\sqrt{E_1}}{N_f}\sum_{j=iN_f}^{(i+1)N_f-1}\phi_w(\epsilon_j^{(1)})+\frac{\sqrt{E_2}}{N_c}\gamma^{(2)}_1}{\sqrt{\frac{E_2}{N}[\gamma^{(2)}_2-(\gamma^{(2)}_1)^2/N_c]+\sigma_n^2}}\right)
+\,\frac{1}{2}\,Q\left(\frac{\frac{\sqrt{E_1}}{N_f}\sum_{j=iN_f}^{(i+1)N_f-1}\phi_w(\epsilon_j^{(1)})-\frac{\sqrt{E_2}}{N_c}\gamma^{(2)}_1}{\sqrt{\frac{E_2}{N}[\gamma^{(2)}_2-(\gamma^{(2)}_1)^2/N_c]+\sigma_n^2}}\right).
\end{gather}

Then, for large $N_f$ values, we can again invoke the CLT for
$\frac{1}{\sqrt{N_f}}\sum_{j=iN_f}^{(i+1)N_f-1}[\phi_w(\epsilon_j^{(1)})-\mu]$
and approximate the unconditional BEP as
\begin{gather}
P_e\approx\frac{1}{2}Q\left(\frac{\sqrt{E_1}\mu+\frac{\sqrt{E_2}}{N_c}\gamma_1^{(2)}}{\sqrt{\frac{E_1\sigma^2}{N}N_c+\frac{E_2}{N}[\gamma^{(2)}_2-(\gamma^{(2)}_1)^2/N_c]+\sigma_n^2}}\right)
+\,\frac{1}{2}Q\left(\frac{\sqrt{E_1}\mu-\frac{\sqrt{E_2}}{N_c}\gamma_1^{(2)}}{\sqrt{\frac{E_1\sigma^2}{N}N_c+\frac{E_2}{N}[\gamma^{(2)}_2-(\gamma^{(2)}_1)^2/N_c]+\sigma_n^2}}\right).
\end{gather}

For the multiuser case, assume that all the interfering users have
the same energy $E$ and probability distributions of the jitters
are i.i.d. for all of them. Then, the total MAI can be
approximated by a zero mean Gaussian random variable
for sufficiently large number of users, $N_u$, and, after similar
manipulations, the BEP can be expressed approximately as
\begin{gather}\label{eq:BEP_uncoded}
P_e\approx
Q\left(\frac{\sqrt{E_1}\mu}{\sqrt{\frac{E_1\sigma^2}{N}N_c+(N_u-1)E\left(\frac{\gamma_2}{N}+\frac{\gamma_1^2}{N_c^2}-\frac{\gamma_1^2}{NN_c}\right)+\sigma_n^2}}\right),
\end{gather}
where the user index $k$ is dropped from $\gamma_1^{(k)}$ and
$\gamma_2^{(k)}$ since they are i.i.d. among interfering users.

Considering (\ref{eq:BEP_uncoded}), it is seen that for relatively
small $N_c$ values, the second term in the denominator, which is
the term due to MAI, can become large and cause an increase in the
BEP. Similarly, when $N_c$ is large, the first term in the
denominator can become significant and the BEP can become high
again. Therefore, we expect to have an optimal $N_c$ value for the
interference-limited case. Intuitively, for small $N_c$ values,
the number of pulses per bit, $N_f$, is large. Therefore, we have
high BEP due to large amount of MAI. As $N_c$ becomes large, the
MAI becomes more negligible. However, making $N_c$ very large can
again cause an increase in BEP since $N_f$ becomes small, in which
case the effect of timing jitter becomes more significant. The
optimal $N_c$ ($N_f$) value can be approximated by using
(\ref{eq:BEP_uncoded}).

Now consider the chip-synchronous case. In this case, the
following lemma approximates the distribution of the overall MAI
for large number of equal energy interferers:
\begin{lemma:uncoded_chip}\label{lem:uncoded_chip}
Let $N\longrightarrow \infty$ and $\frac{N_c}{N_f}\longrightarrow
c>0$. Assume that all $(N_u-1)$ interfering users have the same
bit energy $E$ and i.i.d. jitter statistics. Then, the overall
MAI, $a$ in (\ref{eq:MF_out}), is approximately distributed, for
large $N_u$, as
\begin{gather}\label{eq:lemma_asyc}
a \sim
{\mathcal{N}}\left(0\,,\,\frac{E(N_u-1)}{N_c}\left[\gamma_2+\frac{(N_f-1)[2N_c^2(N_f-1)+1]}{3NN_c^2}\gamma_1^2\right]\right),
\end{gather}
where $\gamma_1$ and $\gamma_2$ are as in (\ref{eq:parameters}).
\end{lemma:uncoded_chip}

\textit{Proof:} The proof is omitted due to space limitations. It
mainly depends on some central limit arguments.

Comparing the variance in Lemma \ref{lem:uncoded_chip} with the
variance of the MAI term in the uncoded symbol-synchronous case
for large number of equal energy interferers with the same jitter
statistics, it can be shown that $\sigma^2_{MAI,\,chip}\leq
\sigma^2_{MAI,\,symb}$ where
\begin{align}
\sigma^2_{MAI,\,symb}&=\frac{E(N_u-1)}{N_c}\left[\gamma_2+\frac{N_f-1}{N_c}\,\gamma_1^2\right],\\
\sigma^2_{MAI,\,chip}&=\frac{E(N_u-1)}{N_c}\left[\gamma_2+\frac{(N_f-1)[2N_c^2(N_f-1)+1]}{3NN_c^2}\,\gamma_1^2\right],
\end{align}
where the equality is satisfied only for $N_f=1$.

The reason behind this inequality can be explained as follows: In
the uncoded symbol-synchronous case, interference components from
a given user to the pulses of the template signal has the same
polarity and therefore they add coherently for each user. However,
in the chip-synchronous case, interference to some pulses of the
template is due to one information bit whereas the interference to
the remaining pulses is due to another information bit because
there is a misalignment between symbol transmission instants
(Figure \ref{fig:Async}). Since information bits can be $\pm1$
with equal probability, the interference from a given user to
individual pulses of the template signal does not always add
coherently. Therefore, the average power of the MAI is smaller in
the chip-synchronous case. As the limiting case, consider the
coded case, where each individual pulse has a random polarity
code. In this case, the overall interference from a user, given
the information bit of that user, is zero mean due to the polarity
codes. Hence, the overall interference from all users has a
smaller average power, given by $\gamma_2E(N_u-1)/N_c$, for equal
energy interferers with i.i.d jitter statistics.

By Lemma \ref{lem:uncoded_chip} and the approximation to the
distribution of the signal part of the MF output given the
information bit in (\ref{eq:MF_out}) by a Gaussian random
variable, we get
\begin{gather}\label{eq:BEP_uncoded_asyc}
P_e\approx
Q\left(\frac{\sqrt{E_1}\mu}{\sqrt{\frac{E_1\sigma^2}{N}N_c+(N_u-1)E\left[\frac{\gamma_2}{N}+\frac{(N-N_c)[2N_c(N-N_c)+1]}{3N^2N_c^3}\gamma_1^2\right]+\sigma_n^2}}\right),
\end{gather}
where $\mu=\textrm{E}\{\phi_w(\epsilon_j^{(1)})\}$ and
$\sigma^2=\textrm{Var}\{\phi_w(\epsilon_j^{(1)})\}$.

Considering (\ref{eq:BEP_uncoded_asyc}), we have similar
observations as in the symbol-synchronous case. Considering the
interference-limited case, for small $N_c$ values, the second term
in the denominator, which is the term due to the MAI, becomes
dominant and causes a large BEP. When $N_c$ is large, the MAI
becomes less significant since probability of overlaps between
pulses decreases. However, for very small $N_c$ values, the effect
of the timing jitter can become more significant as can be seen
from the first term in the denominator and the BEP can increase
again. Therefore, in this case, we again expect to see a trade-off
between processing gain parameters.

\section{Multipath Case}\label{sec:MP}

In this section, the effects of the processing gain parameters,
$N_c$ and $N_f$, on the BEP performance of a coded system are
investigated in a frequency-selective environment. The following
channel model is considered \cite{ieee802153a}, \cite{Win_charac}:
\begin{gather}\label{eq:channel_MP}
h(t)=\sum_{l=0}^{L-1}\alpha_l\delta(t-\tau_l),
\end{gather}
where $\alpha_l$ and $\tau_l$ are, respectively, the fading
coefficient and the delay of the $(l+1)$th path. Note that we
assume a single cluster without loss of generality of the
analysis. In fact, even a multipath channel model with pulse
distortions can be incorporated into the analysis as will be
explained at the end of the section.

We consider a downlink scenario, where the transmitted symbols are
synchronized, and assume $\tau_0=0$ without loss of generality.
Moreover, for the simplicity of the analysis, the delay of the last
path, $\tau_{L-1}$, is set to an integer multiple of the chip
interval $T_c$; that is, $\tau_{L-1}=(M-1)T_c$ where $M$ is an
integer. Note that this does not cause a loss in generality since we
can always think of a hypothetical path at
$T_c\lceil\tau_{L-1}/T_c\rceil$ with a fading coefficient of zero,
with $\lceil x\rceil$ denoting the smallest integer larger than or
equal to $x$.

From (\ref{eq:tran1}) and (\ref{eq:channel_MP}), the received
signal can be expressed as follows:
\begin{gather}\label{eq:recSig_MP}
r(t)=\sum_{k=1}^{N_u}\sqrt{\frac{E_k}{N_f}}\sum_{j=-\infty}^{\infty}d_j^{(k)}b^{(k)}_{\lfloor
j/N_f\rfloor}u(t-jT_f-c_j^{(k)}T_c-\epsilon_j^{(k)})+\sigma_nn(t),
\end{gather}
where $n(t)$ is zero mean white Gaussian noise with unit spectral
density and $u(t)=\sum_{l=0}^{L-1}\alpha_lw_{{\rm{rx}}}(t-\tau_l)$,
with $w_{{\rm{rx}}}(t)$ denoting the received unit energy UWB pulse.
$\epsilon_j^{(k)}$ is the timing jitter at the transmitted pulse in
the $j$th frame of user $k$. We assume that $\epsilon^{(k)}_j$ for
$j=\ldots,-1,0,1,\ldots$ form an i.i.d. sequence for each user and
that $\underset{j,k}{\max}\,|\epsilon_j^{(k)}|<T_c$.

Consider a generic Rake receiver that combines a number of multipath
components of the incoming signal. Rake receivers are considered for
UWB systems in order to collect sufficient signal energy from
incoming multipath components
\cite{Cassioli_2002_UWBrake}-\cite{Win_2000_Perf RAKE}. For the
$i$th information symbol, the following signal represents the
template signal for such a receiver:
\begin{gather}\label{eq:template_MP}
s_{{\rm{temp}}}^{(1)}(t)=\sum_{j=iN_f}^{(i+1)N_f-1}d_j^{(1)}v_j(t-jT_f-c_j^{(1)}T_c),
\end{gather}
where user $1$ is considered as the user of interest without loss
of generality,
$v_j(t)=\sum_{l=0}^{L-1}\beta_lw_{{\rm{rx}}}(t-\tau_l-\hat{\epsilon}_{j,l})$,
with $\boldsymbol{\beta}=[\beta_{0}\cdots\beta_{L-1}]$ denoting
the Rake combining coefficients and $\hat{\epsilon}_{j,l}$ being
the timing jitter at the $l$th finger in the $j$th frame of the
template signal. We assume that
$\underset{j,l}{\max}\,|\hat{\epsilon}_{j,l}|<T_c$ and
$\underset{i,j,k,l}{\max}\,|\hat{\epsilon}_{j,l}-\epsilon_i^{(k)}|<T_c$,
which, practically true most of the time, makes sure that a pulse
can only interfere with the neighboring chip positions due to
timing jitter.

Corresponding to different situations, we consider three different
statistics for the jitter at the template signal:

\textit{Case-1:} The jitter is assumed to be i.i.d. for all
different finger and frame indices; that is,
$\hat{\epsilon}_{j,l}$ for $(j,l)\in\mathcal{Z}\times\mathcal{L}$
form an i.i.d. sequence, where where $\mathcal{Z}$ is the set of
integers and $\mathcal{L}=\{0,1,\ldots,L-1\}$.

\textit{Case-2:} The same jitter value is assumed for all fingers,
and i.i.d. jitters are assumed among different frames. In other
words, $\hat{\epsilon}_{j,l_1}=\hat{\epsilon}_{j,l_2}$ $\forall
l_1,l_2$ and $\hat{\epsilon}_{j,l}$ for $j\in\mathcal{Z}$ form an
i.i.d. sequence.

\textit{Case-3:} The same jitter value is assumed in all frames
for a given finger, and i.i.d. jitters are assumed among the
fingers. In other words,
$\hat{\epsilon}_{j_1,l}=\hat{\epsilon}_{j_2,l}$ $\forall j_1,j_2$
and $\hat{\epsilon}_{j,l}$ for $l\in{\mathcal{L}}$ form an i.i.d.
sequence.

We will consider only Case-1 and Case-2 in the following analysis
and an extension of the results to Case-3 will be briefly
discussed at the end of the section.

Using (\ref{eq:recSig_MP}) and (\ref{eq:template_MP}), the
correlation output for the $i$th symbol can be expressed as
\begin{gather}\label{eq:cor_out_MP}
y=\int
r(t)\,s_{{\rm{temp}}}^{(1)}(t)\,dt=b_i^{(1)}\sqrt{\frac{E_1}{N_f}}\sum_{m=iN_f}^{(i+1)N_f-1}\phi_{uv}^{(m)}(\epsilon_m^{(1)})+\hat{a}+a+n,
\end{gather}
where the first term is the desired signal component with
$\phi_{uv}^{(m)}(\Delta)=\int u(t-\Delta)v_m(t)dt$, $a$ is the MAI,
$\hat{a}$ is the inter-frame interference (IFI) and
$n=\sigma_n\sum_{j=iN_f}^{(i+1)N_f-1}d_j^{(1)}\int
v_j(t-jT_f-c_j^{(1)}T_c)n(t)dt$ is the output noise, which can be
shown to be distributed, approximately, as
$n\sim\mathcal{N}\left(0\,,\,\sigma_n^2N_{{\rm{f}}}\bar{E}_v\right)$,
with
$\bar{E}_v=\frac{1}{N_f}\sum_{j=iN_f}^{(i+1)N_f-1}\int_{-\infty}^{\infty}v_j^2(t)dt$,
for large $N_{{\rm{f}}}$${}^8$\footnotetext[8]{$\bar{E}_v$ is
approximately independent of $N_f$ in most practical cases.}.

The IFI is the self interference among the pulses of the user of
interest, user $1$, which occurs when a pulse in a frame spills over
to adjacent frame(s) due to multipath and/or timing jitter and
interferes with a pulse in that frame. The overall IFI can be
considered as the sum of interference to each frame; that is,
$\hat{a}={\sqrt\frac{E_1}{N_f}}\sum_{m=iN_f}^{(i+1)N_f-1}\hat{a}_m$,
where the interference to the $m$th pulse can be expressed as
\begin{gather}\label{eq:a_m_hat}
\hat{a}_m=d_m^{(1)}\int v_m(t-mT_f-c_m^{(1)}T_c)\underset{j\ne
m}{\sum_{j=-\infty}^{\infty}}d^{(1)}_j\, b^{(1)}_{\lfloor
j/N_f\rfloor}u(t-jT_f-c^{(1)}_jT_c-\epsilon_j^{(1)})dt.
\end{gather}

Assume that the delay spread of the channel is not larger than the
frame time. In other words, $M\leq N_c$. In this case,
(\ref{eq:a_m_hat}) can be expressed as
\begin{gather}\label{eq:a_m_hat_2}
\hat{a}_m=d_m^{(1)}\sum_{i\in\{-1,1\}}d_{m+i}^{(1)}b^{(1)}_{\lfloor(m+i)/N_f\rfloor}\phi_{uv}^{(m)}\left(iT_f+(c_{m+i}^{(1)}-c_{m}^{(1)})T_c+\epsilon_{m+i}^{(1)}\right).
\end{gather}

Then, using the central limit argument in \cite{meas} for
dependent sequences, we can obtain the distribution of $\hat{a}$
as follows:
\begin{lemma:IFI}\label{lem:IFI}
As $N\longrightarrow \infty$ and $\frac{N_f}{N_c}\longrightarrow
c>0$, the IFI $\hat{a}$ is asymptotically normally distributed as
\begin{gather}\label{eq:lemmaIFI}
\hat{a}\sim
{\mathcal{N}}\left(0\,,\,\frac{E_1}{N_c^2}\sum_{j=1}^{M}j\,\textrm{E}\{[\phi_{uv}^{(m)}(jT_c+\epsilon_{m+1}^{(1)})+\phi_{uv}^{(m)}(-jT_c+\epsilon_{m-1}^{(1)})]^2\}\right).
\end{gather}
\end{lemma:IFI}

\textit{Proof:} See Appendix \ref{app:IFI}.

Note that the result is true for both Case-1 and Case-2. The only
difference between the two cases is the set of jitter variables over
which the expectation is taken.

The MAI term in (\ref{eq:cor_out_MP}) can be expressed as the sum
of interference from each user,
$a=\sum_{k=2}^{N_u}\sqrt{\frac{E_k}{N_f}}\,a^{(k)}$, where each
$a^{(k)}$ can be considered as the sum of interference to each
frame of the template signal from the signal of user $k$. That is,
$a^{(k)}=\sum_{m=iN_f}^{(i+1)N_f-1}a_m^{(k)}$, where
\begin{gather}\label{eq:a_m^k_MP}
a_m^{(k)}=d_m^{(1)}\int
v_m(t-mT_f-c_m^{(1)}T_c)\sum_{j=-\infty}^{\infty}d_j^{(k)}b^{(k)}_{\lfloor
j/N_f\rfloor}u(t-jT_f-c^{(k)}_jT_c-\epsilon_j^{(k)})\,dt.
\end{gather}

Assuming $M\leq N_c$, $a_m^{(k)}$ can be expressed as
\begin{gather}\label{eq:a_m_k_MP_2}
a_m^{(k)}=d_m^{(1)}\sum_{j=m-1}^{m+1}d_j^{(k)}b^{(k)}_{\lfloor
j/N_f\rfloor}\phi_{uv}^{(m)}\left((j-m)T_f+(c_j^{(k)}-c_m^{(1)})T_c+\epsilon_j^{(k)}\right).
\end{gather}

Then, using the same central limit argument \cite{meas} as in
Lemma \ref{lem:IFI}, we obtain the following result:

\begin{lemma:MAI_MP}\label{lem:MAI_MP}
As $N\longrightarrow \infty$ and $\frac{N_f}{N_c}\longrightarrow
c>0$, the MAI from user $k$, $a^{(k)}$, is asymptotically normally
distributed as
\begin{gather}\label{eq:lemmaMAI_MP}
a^{(k)} \sim
{\mathcal{N}}\left(0\,,\,\frac{N_f}{N_c}\sum_{j=-M}^{M}\textrm{E}\{[\phi_{uv}^{(m)}(jT_c+\epsilon^{(k)})]^2\}\right).
\end{gather}
\end{lemma:MAI_MP}

\textit{Proof:} See Appendix \ref{app:MAI_MP}.

Since we assume that the timing jitter variables at the transmitted
pulses in different frames form an i.i.d. sequence for a given user,
and the jitter at the template is i.i.d. among different frames, the
first term in (\ref{eq:cor_out_MP}) given the information bit of
user $1$ converges to a Gaussian random variable for large $N_f$
values. Using this observation and the results of the previous two
lemmas, we can express the BEP of the system as
\begin{gather}\label{eq:BEP_MP}
P_e\approx
Q\left(\frac{\sqrt{E_1}\,\textrm{E}\{\phi_{uv}^{(m)}(\epsilon^{(1)})\}}{\sqrt{\frac{E_1N_c}{N}\textrm{Var}\{\phi_{uv}^{(m)}(\epsilon^{(1)})\}
+\frac{E_1}{N_cN}\sigma_{IFI}^2
+\frac{1}{N}\sum_{k=2}^{N_u}E_k\sigma_{MAI,k}^2+\bar{E}_v\sigma_n^2}}\right),
\end{gather}
where
\begin{gather}
\sigma_{IFI}^2=\sum_{j=1}^{M}j\,\textrm{E}\{[\phi_{uv}^{(m)}(jT_c+\epsilon_{m+1}^{(1)})+\phi_{uv}^{(m)}(-jT_c+\epsilon_{m-1}^{(1)})]^2\}
\end{gather}
and
\begin{gather}\label{eq:SON}
\sigma_{MAI,k}^2=\sum_{j=-M}^{M}\textrm{E}\{[\phi_{uv}^{(m)}(jT_c+\epsilon^{(k)})]^2\}
\end{gather}
are independent of the processing gain parameters.

From (\ref{eq:BEP_MP}), a trade-off between the effect of the
timing jitter and that of the IFI is observed. The first term in
the denominator, which is due to the effect of the timing jitter
on the desired signal part of the output in (\ref{eq:cor_out_MP}),
can cause an increase in the BEP as $N_c$ increases. The second
term in the denominator is due to the IFI, which can cause a
decrease in the BEP as $N_c$ increases; because, as $N_c$
increases, the probability of a spill-over from one frame to the
next decreases. Hence, large $N_c$ values mitigate the effects of
IFI. The term due to the MAI (the third term in the denominator)
does not depend on $N_c$ ($N_f$) for a given value of total
processing gain $N$. Therefore, it has no effect on the trade-off
between processing gains. The optimal value of $N_c$ ($N_f$)
minimizes the BEP by optimally mitigating the opposing effects of
the timing jitter and the IFI.

\textbf{Remark-1:} The same conclusions hold for Case-3, in which
the timing jitter at the template signal is the same for all frames
for a given finger, and i.i.d. among different fingers. In this
case, conditioning on the jitter values at different fingers
($\hat{\epsilon}_{j,0}\cdots\hat{\epsilon}_{j,L-1}$), the
conditional BEP
($P_{{\rm{e}}}|\hat{\epsilon}_{j,0}\cdots\hat{\epsilon}_{j,L-1}$)
can be shown to be as in (\ref{eq:BEP_MP}); hence, the same
dependence structure on the processing gain parameters is observed.
The only difference in this case is that the statistical averages
are calculated only over the jitter values at the transmitter.

\textbf{Remark-2:} For the case in which pulses are also distorted
by the channel; i.e., pulse shapes in different multipath components
are different, the analysis is still valid. Since the results
(\ref{eq:BEP_MP})-(\ref{eq:SON}) are in terms of the
cross-correlation $\phi_{uv}^{(j)}(\cdot)$ of
$u(t)=\sum_{l=0}^{L-1}\alpha_lw_{{\rm{rx}}}(t-\tau_l)$ and
$v_j(t)=\sum_{l=0}^{L-1}\beta_lw_{{\rm{rx}}}(t-\tau_l-\hat{\epsilon}_{j,l})$,
by replacing these expressions by
$u(t)=\sum_{l=0}^{L-1}\alpha_lw^{(l)}_{{\rm{rx}}}(t-\tau_l)$ and
$v_j(t)=\sum_{l=0}^{L-1}\beta_lw^{(l)}_{{\rm{rx}}}(t-\tau_l-\hat{\epsilon}_{j,l})$,
where $w^{(l)}_{{\rm{rx}}}(t)$ represents the received pulse from
the $(l+1)$th signal path, generalizes the analysis to the case in
which the channel introduces pulse distortions.

\section{Simulation Results}

\begin{figure}
\begin{center}
\includegraphics[width = 0.75\textwidth]{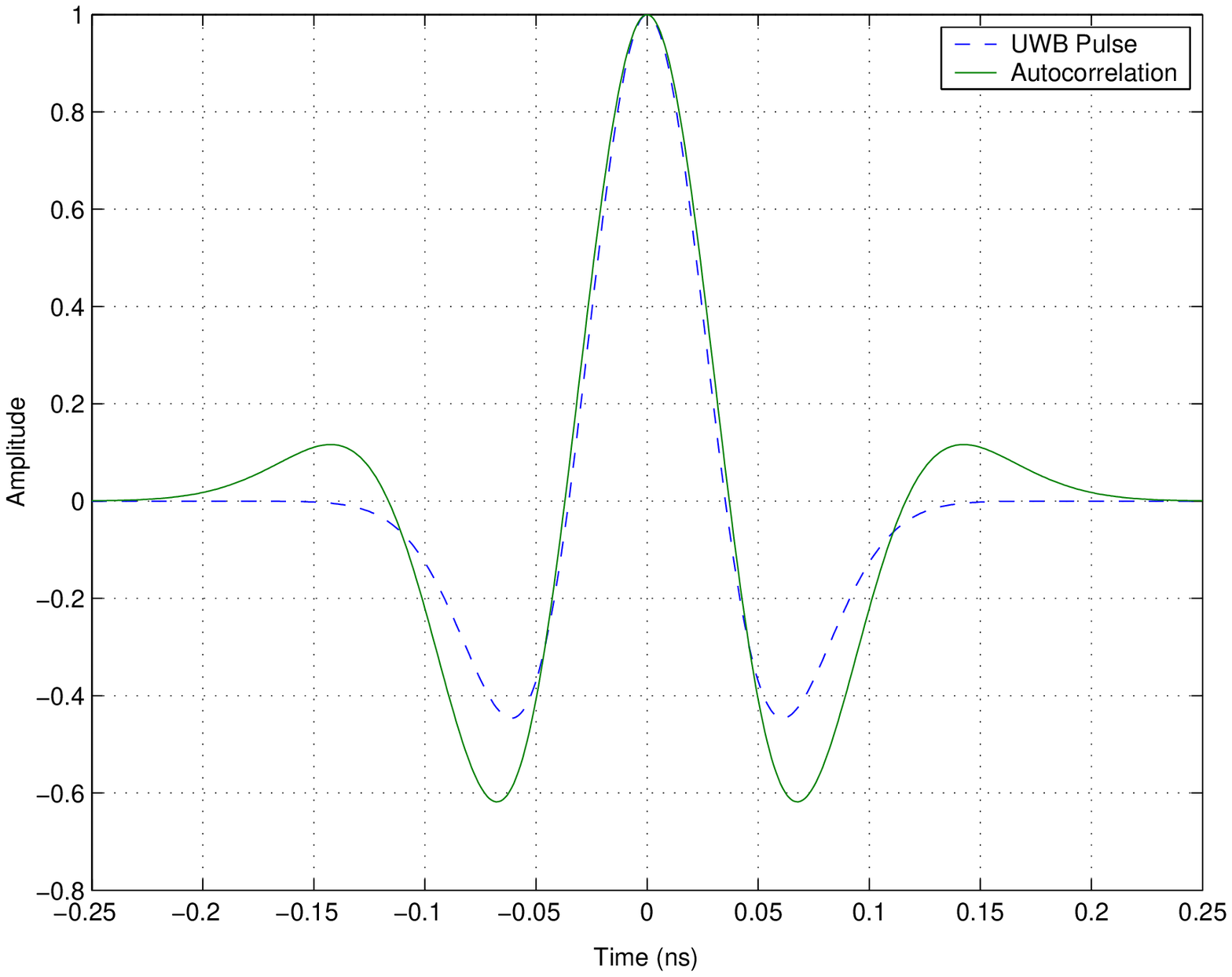}
\caption{The UWB pulse and the autocorrelation function for
$T_c=0.25\,$ns.} \label{fig:pulse}
\end{center}
\end{figure}

In this section, BEP performances of coded and uncoded IR systems
are simulated for different values of processing gains, and the
results are compared with the theoretical analysis. The UWB
pulse$^{9}$\footnotetext[9]{$w_{{\rm{rx}}}(t)=w(t)/\sqrt{E_p}$ with
$E_p=\int_{-\infty}^{\infty}w^2(t)dt$ is used as the received UWB
pulse with unit energy.} and the normalized autocorrelation function
used in the simulations are as follows \cite{ramirez}:
\begin{gather}
w(t)=\left(1-\frac{4\pi t^2}{\tau^2}\right)e^{-2\pi
t^2/\tau^2},\quad\,\, R(\Delta t)=\left[1-4\pi(\frac{\Delta
t}{\tau})^2+\frac{4\pi^2}{3}(\frac{\Delta
t}{\tau})^4\right]e^{-\pi (\frac{\Delta t}{\tau})^2},
\end{gather}
where $\tau=0.125\,\rm{ns}$ is used (Figure \ref{fig:pulse}).


For the first set of simulations, the timing jitter at the
transmitter is modelled by
$\mathcal{U}[-25\,{\rm{ps}},25\,{\rm{ps}}]$, where
$\mathcal{U}[x,y]$ denotes the uniform distribution on $[x,y]$
\cite{Ismail_jitter}, \cite{2005_Zhang_Jitter resist}, and $T_c$
is chosen to be $0.25\,{\rm{ns}}$. The total processing gain
$N=N_cN_f$ is taken to be $512$. Also all $10$ users ($N_u=10$)
are assumed to be sending unit-energy bits ($E_k=1$ $\forall k$)
and $\sigma_n^2=0.1$.

\begin{figure}
\begin{center}
\includegraphics[width = 0.80\textwidth]{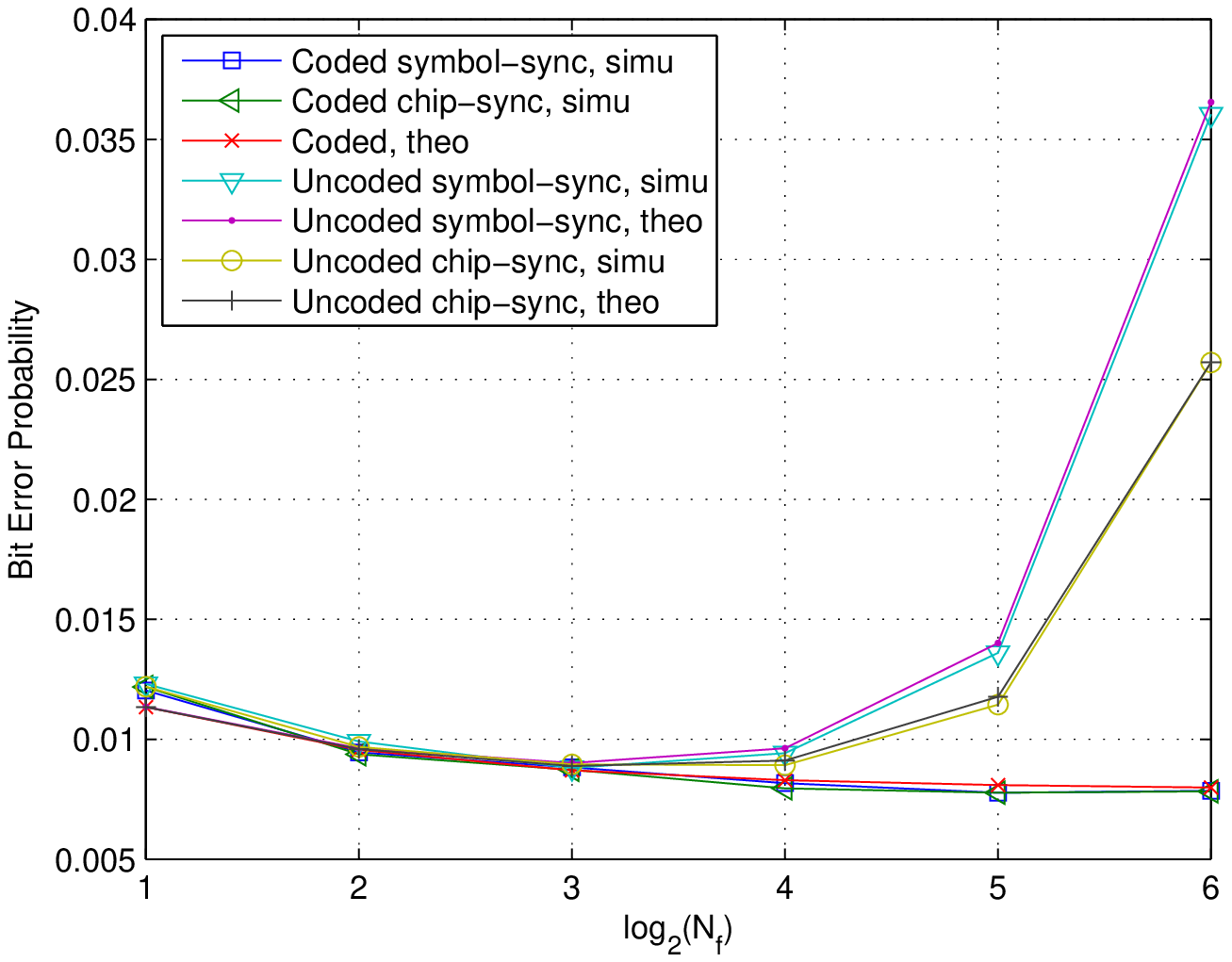}
\caption{BEP vs $\log_2N_{{\rm{f}}}$ for coded and uncoded IR-UWB
systems for the AWGN channel case.} \label{fig:both}
\end{center}
\end{figure}

Figure \ref{fig:both} shows the BEP of the coded and the uncoded
IR-UWB systems for different $N_f$ values in an AWGN environment. It
is observed that the simulation results match quite closely with the
theoretical values.
For the coded system, the BEP decreases as $N_f$ increases. Since
the effect of the MAI on the BEP is asymptotically independent of
$N_f$, the only effect to consider is that of the timing jitter.
Since the effect of the timing jitter is reduced for large $N_f$,
the plots for the coded system show a decrease in BEP as $N_f$
increases. As expected, the performance is the same for the
symbol-synchronous and chip-synchronous coded systems. For the
uncoded system, there is an optimal value of the processing gain
that minimizes the BEP of the system. In this case, there are both
the effects of the timing jitter and the MAI. The effect of the
timing jitter is mitigated using large $N_f$, while that of the
MAI is reduced using small $N_f$. The optimal value of the
processing gains can be approximately calculated using
(\ref{eq:BEP_uncoded}) or (\ref{eq:BEP_uncoded_asyc}). As
expected, the effect of the MAI is larger for the
symbol-synchronous system, which causes a larger BEP for such
systems compared to the chip-synchronous ones.

\begin{figure}
\begin{center}
\includegraphics[width = 0.75\textwidth]{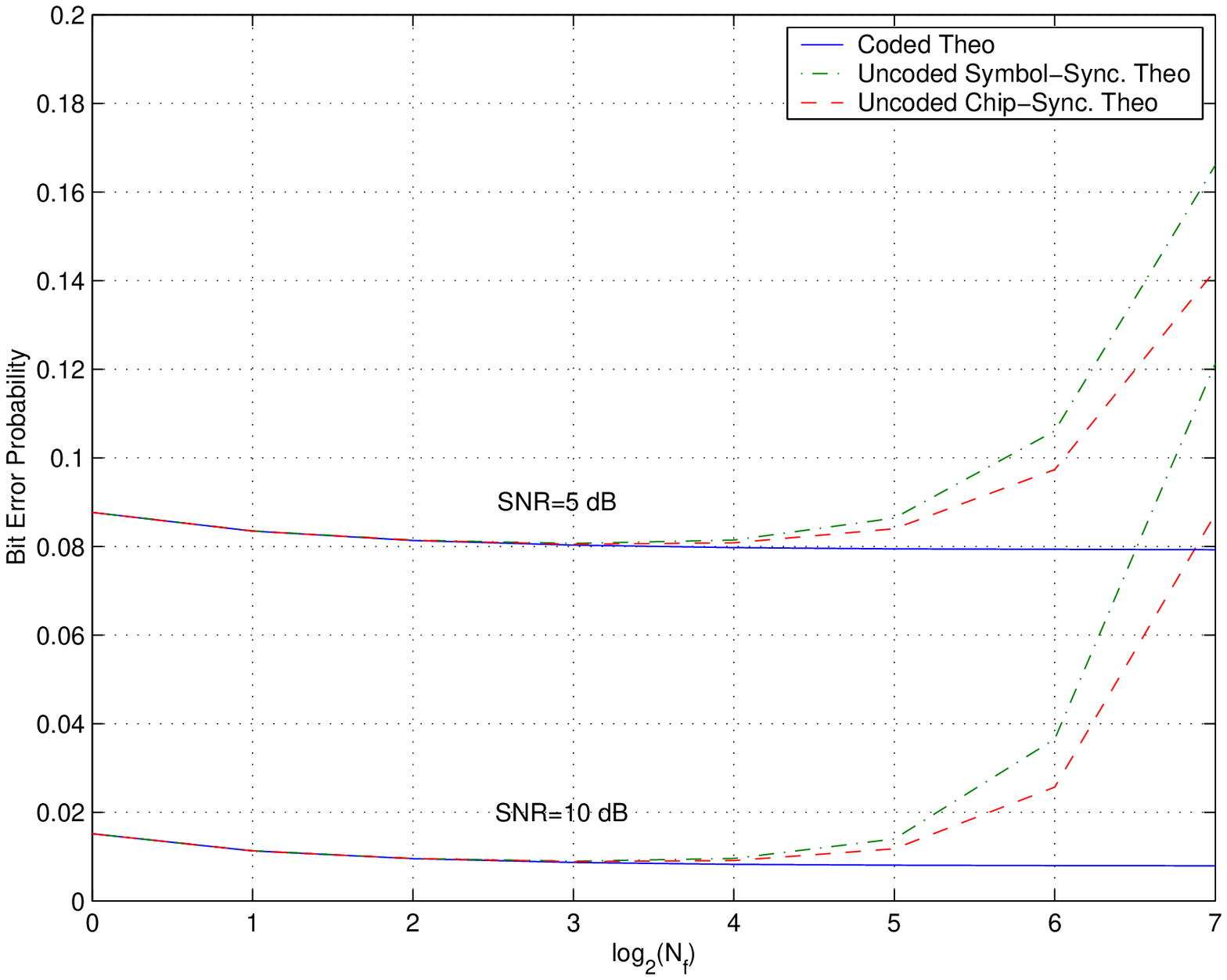}
\caption{Theoretical BEP vs $\log_2N_{{\rm{f}}}$ curves for coded
and uncoded IR-UWB systems for different SNR values.}
\label{fig:diffSNR}
\end{center}
\end{figure}

In Figure \ref{fig:diffSNR}, the BEPs of the coded and uncoded
systems are plotted for different signal-to-noise ratios (SNRs).
As observed from the figure, as the SNR increases, the trade-off
between the processing gains become more significant. This is
because, for large SNR, the background noise gets small compared
to the noise due to jitter or MAI; hence, the change of processing
gain parameters causes significant changes in the BEP of the
system due to the effects of timing jitter (and MAI in the uncoded
case).

\begin{figure}
\begin{center}
\includegraphics[width = 0.75\textwidth]{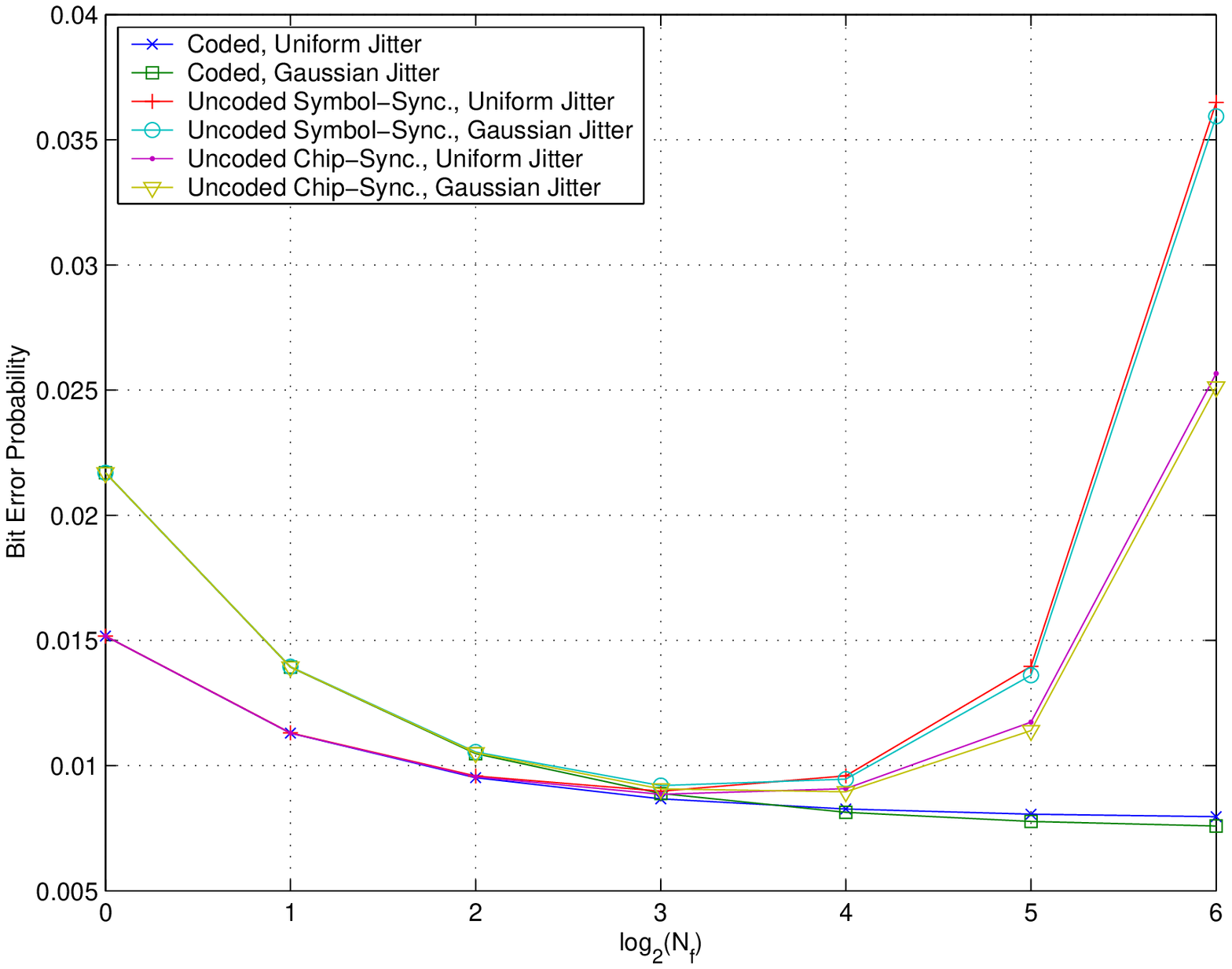}
\caption{Theoretical BEP vs $\log_2N_{{\rm{f}}}$ curves for coded
and uncoded IR-UWB systems for uniform and Gaussian jitter
statistics.} \label{fig:diffJits}
\end{center}
\end{figure}

In Figure \ref{fig:diffJits}, the effects of jitter distribution
on the system performance are investigated. The BEPs of the coded
and uncoded systems are plotted for zero mean uniform
\cite{2005_Zhang_Jitter resist} and Gaussian
\cite{2006_Kokalis_MUI jitter} timing jitters with the same
variance ($208.3$ ps${}^2$). From the figure, it is observed that
the Gaussian timing jitter increases the BEP more than the uniform
timing jitter; i.e. as the timing jitter becomes significant (for
small $N_f$), the BEP of the system with Gaussian jitter gets
larger than that of the system with uniform jitter. The main
reason for this difference is that the Gaussian jitter can take
significantly large values corresponding to the tail of the
distribution, which considerably affects the average BEP of the
system.

Now consider a multipath channel with $L=10$ paths, where the
fading coefficients are given by
$[0.4653\,\,\,\,0.5817\,\,\,\,0.2327
-0.4536\,\,\,\,0.3490\,\,\,\,0.2217 -0.1163\,\,\,\,0.0233 -0.0116
-0.0023]$, and the delays by $\tau_l=lT_c$ for $l=0,\ldots,L-1$,
where $T_c=0.25\,$ns. The Rake receiver combines all the multipath
components using maximal ratio combining (MRC). The jitter is
modelled as $\mathcal{U}[-20\,{\rm{ps}},20\,{\rm{ps}}]$, the total
processing gain $N$ is equal to $512$, and $\sigma_n^2=0.01$.
There are $10$ users in the system where the first user transmits
bits with unit energy ($E_1=1$), while the others have $E_k = E =
5$ $\forall k$.

\begin{figure}
\begin{center}
\includegraphics[width = 0.85\textwidth]{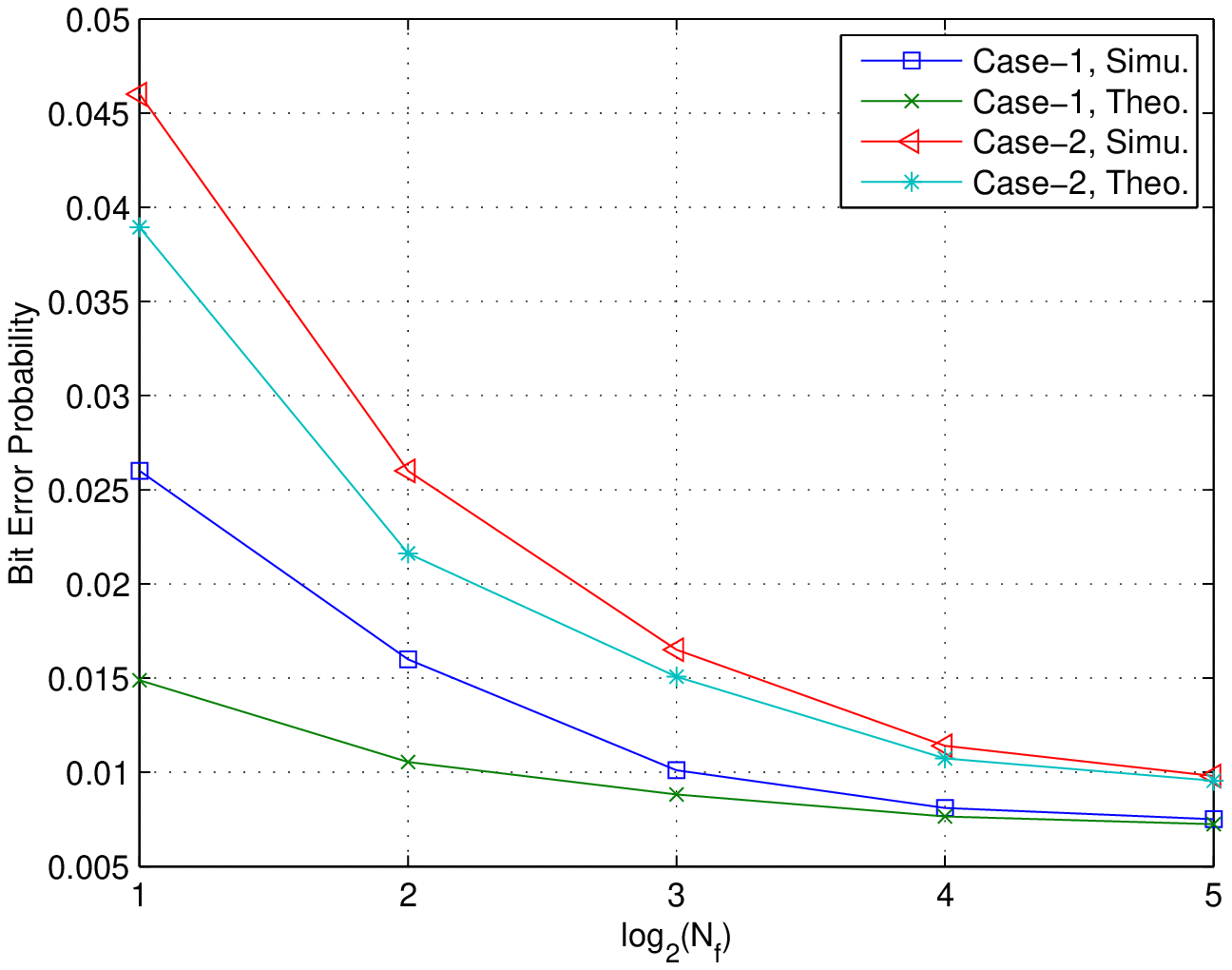}
\caption{BEP versus $N_f$ for coded IR-UWB systems over the
multipath channel $[0.4653\,\,0.5817\,\,0.2327
-0.4536\,\,0.3490\,\,0.2217 -0.1163\,\,0.0233-0.0116-0.0023]$.}
\label{fig:comp_MP}
\end{center}
\end{figure}

Figure \ref{fig:comp_MP} plots the BEP of a coded IR-UWB system
for a downlink scenario, in which user $1$ is considered as the
user of interest and the jitter at the template is as described in
Case-1 and Case-2 in Section \ref{sec:MP}. Note that the
theoretical and simulation results get closer as the number of
pulses per symbol, $N_{{\rm{f}}}$, increases, as the Gaussian
approximation becomes more and more accurate. It is observed that
the BEP decreases as $N_{{\rm{f}}}$ increases since the effect of
timing jitter is reduced. However, the effect of IFI is not
observed since it is negligible compared to the other noise
sources. Hence, no significant increase in BEP is observed for
larger $N_f$, although the IFI increases. Finally, it is observed
that the BEPs for Case-1 are smaller than those for Case-2. In
other words, the effects of timing jitter are smaller when the
jitter is i.i.d. among all the frames and fingers than when it is
the same for all the fingers in a given frame and i.i.d. among
different frames.

\section{Conclusion}

The trade-off between the processing gains of an IR system has been
investigated in the presence of timing jitter. It has been concluded
that, in an AWGN channel, sending more pulses per bit decreases the
BEP of a coded system since the effect of the MAI on the BEP is
independent of processing gains for a given total processing gain,
and the effect of the i.i.d. timing jitter is reduced by sending
more pulses. The system performs the same whether the users are
symbol-synchronous or chip-synchronous. In an uncoded system, there
is a trade-off between $N_c$ and $N_f$, which reflects the effects
of the timing jitter and the MAI. Optimal processing gains can be
found by using the approximate closed form expressions for the BEP.
It is also concluded that the effect of the MAI is mitigated when
the users are chip-synchronous. Therefore, the BEP of a
chip-synchronous uncoded system is smaller than that of a
symbol-synchronous uncoded system.

For frequency-selective environments, the MAI has no effect on the
trade-off between the processing gains of a symbol-synchronous
coded system. However, the IFI is mitigated for larger values of
$N_c$, hence affects the trade-off between the processing gains.
Again the effect of the timing jitter is mitigated by increasing
$N_f$. Therefore, there is a trade-off between the effects of the
timing jitter and that of the IFI and the optimal $N_f$ ($N_c$)
value can be chosen by using the approximate BEP expression.

Related to the trade-off study in this paper, investigation of the
trade-off between processing gain parameters of a transmitted
reference (TR) UWB system \cite{Win_Quek},
\cite{Hoctor_and_Tomlinson_2002} remains as an open research
problem. Due to the autocorrelation receiver employed in TR UWB
systems, the investigation of receiver output statistics would be
more challenging in that case.


\vspace{12pt} {\bf{Acknowledgment:}} The authors would like to
thank Dr. J. Zhang for her support and encouragement.

\appendix
\section{Appendices}

\subsection{Proof of Lemma \ref{lem:coded_AWGN}}\label{app:coded_AWGN}
Assuming a chip-synchronous system, $a_m^{(k)}$ in
(\ref{eq:MAI_l}) can be expressed as
\begin{gather}\label{eq:app_MAI_l}
a_m^{(k)}=d_m^{(1)}\int
w_{{\rm{rx}}}(t-mT_f-c_m^{(1)}T_c)\sum_{j=-\infty}^{\infty}d_j^{(k)}b^{(k)}_{\lfloor
j/N_f\rfloor}w_{{\rm{rx}}}(t-jT_f-c^{(k)}_jT_c-\epsilon_j^{(k)}-\Delta_2^{(k)}T_c)\,dt,
\end{gather}
where $\Delta_2^{(k)}\in\{0,1,\ldots,N-1\}$ with equal
probability.

Due to random polarity codes $d_j^{(k)}$, the distribution of
$a_m^{(k)}$ is the same for all $\Delta_2^{(k)}$ values having the
same $\Delta^{(k)}_1=\mod\{\Delta_2^{(k)},N_c\}$ value. Hence, it
is sufficient to consider
$\Delta_2^{(k)}=\Delta_1^{(k)}\in\{0,1,...,N_c-1\}$. Then,
(\ref{eq:app_MAI_l}) can be expressed as
\begin{gather}\label{eq:app_a_m_3lu}
a_m^{(k)}=d_m^{(1)}\sum_{j=m-1}^{m+1} d_{j}^{(k)}b_{\lfloor
j/N_f\rfloor}\phi_w\left((j-m)T_f+(c_j^{(k)}-c_{m}^{(1)})T_c+\epsilon_{j}^{(k)}+\Delta_1^{(k)}T_c\right)
\end{gather}

From (\ref{eq:app_a_m_3lu}), it is observed that
$\textrm{E}\{a_m^{(k)}\}=0$ due to the independence of polarity
codes for different frame and user indices. Also considering the
random TH sequences and the polarity codes, it can be shown that
\begin{gather}
\textrm{E}\{(a_m^{(k)})^2|\Delta_1^{(k)}\}=\frac{1}{N_c}\left[\textrm{E}\{\phi_w^2(\epsilon^{(k)})\}
+\textrm{E}\{\phi_w^2(T_c-|\epsilon^{(k)}|)\}\right].
\end{gather}

Note that $\textrm{E}\{(a_m^{(k)})^2|\Delta_1^{(k)}\}$ is
independent of $\Delta_1^{(k)}$, which means that the results is
true for both the symbol-synchronous and chip-synchronous cases.

Note that $a_{iN_f}^{(k)},...,a_{(i+1)N_f-1}^{(k)}$ are
identically distributed but not independent. However, they form a
$1$-dependent sequence \cite{meas}. Therefore, for large $N_f$
values,
$\frac{1}{\sqrt{N_f}}\sum_{m=iN_f}^{(i+1)N_f-1}a_{m}^{(k)}$
converge to a zero mean Gaussian random variable with variance
$\textrm{E}\{(a_{iN_f}^{(k)})^2\}+2\textrm{E}\{a_{iN_f}^{(k)}a_{iN_f+1}^{(k)}\}$
\cite{meas}. It is easy to show that the cross-correlation term is
zero using the independence of polarity codes for different
indices. Hence, for large $N_f$ values, $a^{(k)}$ in
(\ref{eq:MAI_l}) is approximately distributed as
$a^{(k)}\sim\mathcal{N}(0\,,\,\gamma^{(k)}_2N_f/N_c)$, where
$\gamma^{(k)}_2=\textrm{E}\{\phi_w^2(\epsilon^{(k)})\}
+\textrm{E}\{\phi_w^2(T_c-|\epsilon^{(k)}|)\}$.

\subsection{Proof of Lemma \ref{lem:uncoded_sym}}\label{app:uncoded_sym}
Let $p_m^{(k)}$ denote the position of the pulse of user $k$ in
the $m$th frame ($p_m^{(k)}=1,...,N_c$). Note that $p_m^{(1)}$
denotes the position of the pulse of the template signal in the
$m$th frame, assuming user $1$ as the user of interest.

For $p_m^{(1)}=2,...,N_c-1$, there occurs interference from user
$k$ to the $m$th pulse of the template signal if user $k$ has its
$m$th pulse at the same position as the $m$th pulse of the
template signal or it has its $m$th pulse at a neighboring
position to $m$th pulse of the template signal and there is a
partial overlap due to the effect of timing jitter. Then
$a_m^{(k)}$ in (\ref{eq:MAI_l}) can be expressed as
\begin{gather}\label{eq:uncoded_syc_1}
a_m^{(k)}=b_i^{(k)}[\,\phi_w(\epsilon_m^{(k)})\,I_{\{p_m^{(1)}=p_m^{(k)}\}}
+\phi_w(T_c-\epsilon_m^{(k)})\,I_{\{p_m^{(1)}-p_m^{(k)}=1\}}I_{\{\epsilon_m^{(k)}>0\}}
+\phi_w(T_c+\epsilon_m^{(k)})\,I_{\{p_m^{(k)}-p_m^{(1)}=1\}}I_{\{\epsilon_m^{(k)}<0\}}],
\end{gather}
where $I_A$ denotes an indicator function that is equal to one in
$A$ and zero otherwise.

For $p_m^{(1)}=1$, we also consider the interference from the
previous frame of the signal received from user $k$:
\begin{gather}\label{eq:uncoded_syc_2}
a_m^{(k)}=b_i^{(k)}[\phi_w(\epsilon_m^{(k)})I_{\{p_m^{(k)}=1\}}
+\phi_w(T_c+\epsilon_m^{(k)})I_{\{p_m^{(k)}=2\}}I_{\{\epsilon_m^{(k)}<0\}}]
+b_i^{(k)}\phi_w(T_c-\epsilon_{m-1}^{(k)})I_{\{p_{m-1}^{(k)}=N_c\}}I_{\{\epsilon_{m-1}^{(k)}>0\}},
\end{gather}
for $m=iN_f+1,...,(i+1)N_f-1$. Note that for $m=iN_f$, we just
need to replace $b_i^{(k)}$ in the third term by $b_{i-1}^{(k)}$
since the previous bit will be in effect in that case.

Similarly, for $p_m^{(1)}=N_c$,
\begin{gather}\label{eq:uncoded_syc_3}
a_m^{(k)}=b_i^{(k)}[\phi_w(\epsilon_m^{(k)})I_{\{p_m^{(k)}=N_c\}}
+\phi_w(T_c-\epsilon_m^{(k)})I_{\{p_m^{(k)}=N_c-1\}}I_{\{\epsilon_m^{(k)}>0\}}]
+b_i^{(k)}\phi_w(T_c+\epsilon_{m+1}^{(k)})I_{\{p_{m+1}^{(k)}=1\}}I_{\{\epsilon_{m+1}^{(k)}<0\}},
\end{gather}
for $m=iN_f,...,(i+1)N_f-2$. For $m=(i+1)N_f-1$, $b_i^{(k)}$ in
the third term is replaced by $b_{i+1}^{(k)}$.

As can be seen from the previous equations,
$a_{iN_f}^{(k)},...,a_{(i+1)N_f-1}^{(k)}$ are not identically
distributed due to the possible small difference for the edge
values $a_{iN_f}^{(k)}$ and $a_{(i+1)N_f-1}^{(k)}$ as stated after
equations (\ref{eq:uncoded_syc_2}) and (\ref{eq:uncoded_syc_3}).
However, those differences become negligible for large $N_c$
and/or $N_f$. Then, $a_{iN_f}^{(k)},...,a_{(i+1)N_f-1}^{(k)}$ can
be considered as identically distributed. The mean value can be
calculated using
$\textrm{E}\{a_m^{(k)}|b_i^{(k)}\}=\textrm{E}\{\textrm{E}\{a_m^{(k)}|\epsilon_{m-1}^{(k)},\epsilon_{m}^{(k)},\epsilon_{m+1}^{(k)},b_i^{(k)}\}\}$.
From equations (\ref{eq:uncoded_syc_1})-(\ref{eq:uncoded_syc_3}),
we get
\begin{eqnarray}\nonumber
\textrm{E}\{a_m^{(k)}|\epsilon_{m-1}^{(k)},\epsilon_{m}^{(k)},\epsilon_{m+1}^{(k)},b_i^{(k)}\}=\frac{N_c-2}{N_c^2}b_i^{(k)}\,[\phi_w(\epsilon_m^{(k)})+\phi_w(T_c-\epsilon_m^{(k)})\,I_{\{\epsilon_m^{(k)}>0\}}
+\phi_w(T_c+\epsilon_m^{(k)})\,I_{\{\epsilon_m^{(k)}<0\}}]\\\nonumber
+\frac{1}{N_c^2}b_i^{(k)}\,[\phi_w(\epsilon_m^{(k)})+\phi_w(T_c+\epsilon_m^{(k)})I_{\{\epsilon_m^{(k)}<0\}}+\phi_w(T_c-\epsilon_{m-1}^{(k)})I_{\{\epsilon_{m-1}^{(k)}>0\}}]\\
+\frac{1}{N_c^2}b_i^{(k)}\,[\phi_w(\epsilon_m^{(k)})+\phi_w(T_c-\epsilon_m^{(k)})I_{\{\epsilon_m^{(k)}>0\}}
+\phi_w(T_c+\epsilon_{m+1}^{(k)})I_{\{\epsilon_{m+1}^{(k)}<0\}}],
\end{eqnarray}
for $m=iN_f,...,(i+1)N_f-1$. Then, taking expectation with respect
to timing jitters, we get
\begin{gather}\label{eq:app_uncoded_sym_mean}
\textrm{E}\{a_m^{(k)}|b_i^{(k)}\}=b_i^{(k)}\gamma^{(k)}_1/N_c,
\end{gather}
where
$\gamma^{(k)}_1=\textrm{E}\{\phi_w(\epsilon_m^{(k)})\}+\textrm{E}\{\phi_w(T_c-|\epsilon_m^{(k)}|)\}$.

By similar calculations, it can be shown that
\begin{align}\nonumber
\textrm{E}\{(a_m^{(k)})^2|b_i^{(k)}\}&=\frac{\gamma^{(k)}_2}{N_c}+\frac{2}{N_c^3}\textrm{E}\{\phi_w(\epsilon^{(k)})\}\textrm{E}\{\phi_w(T_c-|\epsilon^{(k)}|)\}\\
&+\frac{4}{N_c^3}\int_{-\infty}^{0}\phi_w(T_c+\epsilon^{(k)})p(\epsilon^{(k)})d\epsilon^{(k)}
\int_{0}^{\infty}\phi_w(T_c-\epsilon^{(k)})p(\epsilon^{(k)})d\epsilon^{(k)},
\end{align}
where $p(\epsilon^{(k)})$ is the probability density function of
i.i.d. timing jitters for user $k$ and
$\gamma^{(k)}_2=\textrm{E}\{\phi_w^2(\epsilon_m^{(k)})\}+\textrm{E}\{\phi_w^2(T_c-|\epsilon_m^{(k)}|)\}$.
Note that frame indices are omitted in the last equation since the
results do not depend on them.

The cross-correlations between consecutive values of the
1-dependent sequence $a_{iN_f}^{(k)},...,a_{(i+1)N_f-1}^{(k)}$ can
be obtained as
\begin{align}\label{eq:app_uncoded_sym_cross}\nonumber
\textrm{E}\{a_m^{(k)}a_{m+1}^{(k)}|b_i^{(k)}\}&=(\gamma^{(k)}_1)^2/N_c^2-\frac{1}{N_c^3}\gamma^{(k)}_1\textrm{E}\{\phi_w(T_c-|\epsilon^{(k)}|)\}\\
&+\frac{1}{N_c^3}\textrm{E}\{\phi_w(T_c-|\epsilon^{(k)}|)\phi_w(\epsilon^{(k)})\}+\frac{1}{N_c^4}(\textrm{E}\{\phi_w(T_c-|\epsilon^{(k)}|)\})^2.
\end{align}

Then, invoking the theorem for 1-dependent sequences \cite{meas}
and using
(\ref{eq:app_uncoded_sym_mean})-(\ref{eq:app_uncoded_sym_cross}),
the sum of interferences to each pulse of the template,
$\sum_{m=iN_f}^{(i+1)N_f-1}a_m^{(k)}|b_i^{(k)}$, is approximately
distributed as in (\ref{eq:lem_uncod_sym}), where
$\gamma^{(k)}_1$, $\gamma^{(k)}_2$, $\beta^{(k)}_1$ and
$\beta^{(k)}_2$ are as in (\ref{eq:parameters}).

\subsection{Proof of Lemma \ref{lem:IFI}}\label{app:IFI}

In order to calculate the distribution of
$\hat{a}=\sqrt{\frac{E_1}{N_f}}\sum_{m=iN_f}^{(i+1)N_f-1}\hat{a}_m$,
consider $\hat{a}_m$ given by (\ref{eq:a_m_hat_2}). From
(\ref{eq:a_m_hat_2}), it can be observed that
$\textrm{E}\{\hat{a}_m\}=0$ due to the random polarity codes. To
calculate $\textrm{E}\{\hat{a}_m^2\}$, we first condition on the
timing jitter values:
\begin{align}\nonumber
\textrm{E}\{\hat{a}_m^2|\epsilon^{(1)}_{m-1},\epsilon^{(1)}_{m+1},\hat{\boldsymbol{\epsilon}}_{m}\}&=\frac{1}{N_c^2}\sum_{i=0}^{N_c-1}
\sum_{l=0}^{N_c-1}\left[\phi_{uv}^{(m)}\left((l-i-N_c)T_c+\epsilon_{m-1}^{(1)}\right)\right]^2\\\nonumber
&+\frac{1}{N_c^2}\sum_{i=0}^{N_c-1}\sum_{k=0}^{N_c-1}\left[\phi_{uv}^{(m)}\left((k-i+N_c)T_c+\epsilon_{m+1}^{(1)}\right)\right]^2\\
&=\frac{1}{N_c^2}\sum_{j=1}^{M}j\left\{\left[\phi_{uv}^{(m)}(jT_c+\epsilon_{m+1}^{(1)})\right]^2
+\left[\phi_{uv}^{(m)}(-jT_c+\epsilon_{m-1}^{(1)})\right]^2\right\},
\end{align}
where we use the fact that the time-hopping sequences are
uniformly distributed in $\{0,1,\ldots,N_c-1\}$, and
$\hat{\boldsymbol{\epsilon}}_{m}=[\hat{\epsilon}_{m,0}\cdots\hat{\epsilon}_{m,L-1}]$.
Then, averaging over the distribution of the timing jitters, we
get
\begin{gather}\label{eq:appIFI_2}
\textrm{E}\{\hat{a}_m^2\}=\frac{1}{N_c^2}\sum_{j=1}^{M}j\left\{\textrm{E}\{[\phi_{uv}^{(m)}(jT_c+\epsilon_{m+1}^{(1)})]^2\}
+\textrm{E}\{[\phi_{uv}^{(m)}(-jT_c+\epsilon_{m-1}^{(1)})]^2\}\right\}.
\end{gather}

From (\ref{eq:a_m_hat_2}), it is observed that
$\textrm{E}\{\hat{a}_m\hat{a}_n\}=0$ for $|m-n|>1$ and
$\textrm{E}\{\hat{a}_m\hat{a}_{m+1}\}$ can be expressed as
\begin{gather}\label{eq:appIFI_3}
\textrm{E}\{\hat{a}_m\hat{a}_{m+1}\}=\frac{1}{N_c^2}\sum_{j=1}^{M}j\,\textrm{E}\{\phi_{uv}^{(m)}(jT_c+\epsilon_{m}^{(1)})\phi_{uv}^{(m)}(-jT_c+\epsilon_{m+1}^{(1)})\}.
\end{gather}

Since $\frac{1}{\sqrt{N_f}}\sum_{m=iN_f}^{(i+1)N_f-1}\hat{a}_m$
converges to
$\mathcal{N}\left(0\,,\,\textrm{E}\{\hat{a}_m^2\}+2\textrm{E}\{\hat{a}_m\hat{a}_{m+1}\}\right)$
as $N_f\longrightarrow\infty$ \cite{meas}, we can obtain
(\ref{eq:lemmaIFI}) from (\ref{eq:appIFI_2}) and
(\ref{eq:appIFI_3}).

\subsection{Proof of Lemma \ref{lem:MAI_MP}}\label{app:MAI_MP}

The aim is to calculate the asymptotic distribution of
$a^{(k)}=\sum_{m=iN_f}^{(i+1)N_f-1}a_m^{(k)}$, where $a_m^{(k)}$
is given by (\ref{eq:a_m_k_MP_2}).

From (\ref{eq:a_m_k_MP_2}), it can be observed that
$\textrm{E}\{a_m^{(k)}\}=0$ due to the random polarity codes. In
order to calculate the variance, we first condition on the jitter
values:
\begin{align}\nonumber
\textrm{E}\{(a_m^{(k)})^2&|\epsilon^{(k)}_{m-1},\epsilon^{(k)}_{m},\epsilon^{(k)}_{m+1},\hat{\boldsymbol{\epsilon}}_{m}\}=\frac{1}{N_c^2}\sum_{i=0}^{N_c-1}
\{\sum_{j=0}^{N_c-1}\left[\phi_{uv}^{(m)}\left((j-i-N_c)T_c+\epsilon_{m-1}^{(k)}\right)\right]^2\\
&+\sum_{k=0}^{N_c-1}\left[\phi_{uv}^{(m)}\left((k-i)T_c+\epsilon_{m}^{(k)}\right)\right]^2
+\sum_{l=0}^{N_c-1}\left[\phi_{uv}^{(m)}\left((l-i+N_c)T_c+\epsilon_{m+1}^{(k)}\right)\right]^2\},
\end{align}
where use the fact that the time-hopping sequences are uniformly
distributed in $\{0,1,\ldots,N_c-1\}$ and
$\hat{\boldsymbol{\epsilon}}_{m}=[\hat{\epsilon}_{m,0}\cdots\hat{\epsilon}_{m,L-1}]$.
Then, averaging over jitter statistics, we obtain,
$\textrm{E}\{(a_m^{(k)})^2\}=\frac{1}{N_c}\sum_{j=-M}^{M}
\textrm{E}\{[\phi_{uv}^{(m)}(jT_c+\epsilon^{(k)})]^2\}$.

Also it can be observed that $\textrm{E}\{a_m^{(k)}a_n^{(k)}\}=0$
for $m\ne n$ due to the independence of polarity codes.

All in all,
$\frac{1}{\sqrt{N_f}}\sum_{m=iN_f}^{(i+1)N_f-1}a_m^{(k)}$
converges to
$\frac{1}{N_c}\sum_{j=-M}^{M}\textrm{E}\{[\phi_{uv}^{(m)}(jT_c+\epsilon^{(k)})]^2\}$
as $N_f\longrightarrow\infty$ \cite{meas}, from which the result
of Lemma \ref{lem:MAI_MP} follows.

\newpage

\small

\textbf{Sinan Gezici} received the B.S. degree from Bilkent
University, Turkey in 2001, and the Ph.D. degree in Electrical
Engineering from Princeton University in 2006. From April 2006 to
January 2007, he worked as a Visiting Member of Technical Staff at
Mitsubishi Electric Research Laboratories, Cambridge, MA. Since
February 2007, he has been an Assistant Professor in the Department
of Electrical and Electronics Engineering at Bilkent University.

Dr. Gezici's research interests are in the areas of signal
detection, estimation and optimization theory, and their
applications to wireless communications and localization systems.
Currently, he has a particular interest in ultra-wideband systems
for communications and sensing applications.

${}$

\textbf{Andreas F. Molisch} (S'89, M'95, SM'00, F'05) received the
Dipl. Ing., Dr. techn., and habilitation degrees from the Technical
University Vienna (Austria) in 1990, 1994, and 1999, respectively.
From 1991 to 2000, he was with the TU Vienna, becoming an associate
professor there in 1999.  From 2000-2002, he was with the Wireless
Systems Research Department at AT\&T (Bell) Laboratories Research in
Middletown, NJ. Since then, he has been with Mitsubishi Electric
Research Labs, Cambridge, MA, where he is now a Distinguished Member
of Technical Staff. He is also professor and chairholder for radio
systems at Lund University, Sweden.

Dr. Molisch has done research in the areas of SAW filters, radiative
transfer in atomic vapors, atomic line filters, smart antennas, and
wideband systems. His current research interests are measurement and
modeling of mobile radio channels,  UWB, cooperative communications,
and MIMO systems. Dr. Molisch has authored, co-authored or edited
four books (among them the recent textbook ``Wireless
Communications,'' Wiley-IEEE Press), eleven book chapters, some 95
journal papers, and numerous conference contributions.

Dr. Molisch is an editor of the IEEE Trans. Wireless Comm.,
co-editor of recent special issues on MIMO and smart antennas (in J.
Wireless Comm. Mob. Comp.), and UWB (in IEEE - JSAC). He has been
member of numerous TPCs, vice chair of the TPC of VTC 2005 spring,
general chair of ICUWB 2006, and TPC co-chair of the wireless
symposium of Globecomm 2007. He has participated in the European
research initiatives ``COST 231", ``COST 259", and ``COST273", where
he was chairman of the MIMO channel working group, he was chairman
of the IEEE 802.15.4a channel model standardization group, and is
also chairman of Commission C (signals and systems) of URSI
(International Union of Radio Scientists). Dr. Molisch is a Fellow
of the IEEE and recipient of several awards.

${}$

\textbf{H. Vincent Poor} (S'72, M'77, SM'82, F'77) received the
Ph.D. degree in EECS from Princeton University in 1977.  From 1977
until 1990, he was on the faculty of the University of Illinois at
Urbana-Champaign. Since 1990 he has been on the faculty at
Princeton, where he is the Michael Henry Strater University
Professor of Electrical Engineering and Dean of the School of
Engineering and Applied Science. Dr. Poor's research interests are
in the areas of stochastic analysis, statistical signal processing
and their applications in wireless networks and related fields.
Among his publications in these areas is the recent book MIMO
Wireless Communications (Cambridge University Press, 2007).

Dr. Poor is a member of the National Academy of Engineering and is a
Fellow of the American Academy of Arts and Sciences. He is also a
Fellow of the Institute of Mathematical Statistics, the Optical
Society of America, and other organizations.  In 1990, he served as
President of the IEEE Information Theory Society, and in 2004-2007
he served as the Editor-in-Chief of the IEEE Transactions on
Information Theory. Recent recognition of his work includes a
Guggenheim Fellowship and the IEEE Education Medal.

${}$

\textbf{Hisashi Kobayashi} is the Sherman Fairchild University
Professor of Electrical Engineering and Computer Science at
Princeton University since 1986, when he joined the Princeton
faculty as the Dean of the School of Engineering and Applied Science
(1986-91). From 1967 till 1982 he was with the IBM Research Center
in Yorktown Heights, and from 1982 to 1986 he served as the founding
Director of the IBM Tokyo Research Laboratory. His research
experiences include radar systems, high speed data transmission,
coding for high density digital recording, image compression
algorithms, performance modeling and analysis of computers and
communication systems. His current research activities are on
performance modeling and analysis of high speed networks, wireless
communications and geolocation algorithms, network security, and
teletraffic \& queuing theory. He published ``Modeling and
Analysis'' (Addison-Wesley, 1978), and is authoring a new book with
B. L. Mark entitled ``Probability, Statistics and Stochastic
Modeling: Foundations of System Performance Evaluation'' (Prentice
Hall, 2005). He is a member of the Engineering Academy of Japan, a
Life Fellow of IEEE, and a Fellow of IEICE of Japan. He was the
recipient of the Humboldt Prize from Germany (1979), the IFIP's
Silver Core Award (1981), and two IBM Outstanding Contribution
Awards.  He received his Ph.D. degree (1967) from Princeton
University and BE and ME degrees (1961, 63) from the University of
Tokyo, all in Electrical Engineering.

\normalsize

\newpage


\end{document}